\documentclass[a4paper,superscriptaddress,showkeys,prX,showpacs,twocolumn]{revtex4-1}
\usepackage{amssymb, amsmath, amsfonts,bm}
\usepackage{graphics,xcolor}
\usepackage{pst-all}
\usepackage{hyperref}
\usepackage[mathlines,running,modulo]{lineno}
\usepackage{relsize}
\usepackage[utf8]{inputenc}
\usepackage{IEEEtrantools}
\usepackage{epsfig}
\usepackage[toc]{appendix}

\begin{document}
\title{Theoretical determination of Ising-type transition by using the Self-Consistent Harmonic Approximation}
\author{A. R. Moura}
\email{antoniormoura@ufv.br}
\affiliation{Departamento de F\'{i}sica, Universidade Federal de Vi\c{c}osa, 36570-900, Vi\c{c}osa, Minas Gerais, Brazil}
\date{\today}
\begin{abstract}
Over the years, the Self-Consistent Harmonic Approximation (SCHA) has been successfully utilized to determine
the transition temperature of many different magnetic models, particularly the Berezinskii-Thouless-Kosterlitz
transition in  two-dimensional ferromagnets. More recently, the SCHA has found application in describing ferromagnetic
samples in spintronic experiments. In such a case, the SCHA has proven to be an efficient formalism for representing the
coherent state in the ferromagnetic resonance state. One of the main advantages of using the SCHA is the
quadratic Hamiltonian, which incorporates thermal spin fluctuations through renormalization parameters, keeping
the description simple while providing excellent agreement with experimental data. In this article, we investigate the 
SCHA application in easy-axis magnetic models, a subject that has not been adequately explored to date. We obtain both
semiclassical and quantum approaches of the SCHA for a general anisotropic magnetic model and employ them to determine
various quantities such as the transition temperature, spin-wave energy spectrum, magnetization, and critical exponents. 
To verify the accuracy of the method, we compare the SCHA results with experimental and Monte Carlo simulation data for
many distinct well-known magnetic materials. 
\end{abstract}
  

\maketitle

\section{Introduction and motivation}
\label{sec.intro}
The continuous advancements in material science have driven the device miniaturization to sizes approaching
the atomic scale. In this scale, dimensionality plays a crucial role and, usually, the
three-dimensional Physics of macroscopic models does not apply as one might expect. An example of this is
the Berezinskii-Thouless-Kosterlitz (BKT) transition, defined by a topological phase transition exclusively of
two-dimensional models with order parameters exhibiting continuous $O(2)$ symmetry, typically found 
in magnetic models with easy-plane anisotropy. The BKT transition was initially confirmed in the 
superfluid state of thin Helium films\cite{prl40.1727}, as well as in controlled experiments involving 
ultracold trapping of 2D Bose gas\cite{nature441.7097}. Although the 
theoretical analysis of BKT transition has been performed in magnetic models over the years, obtaining 
experimental results has been challenging due to the difficulty of finding genuine two-dimensional 
magnetic materials with planar anisotropy. In many cases, the systems studied are quasi-two-dimensional magnets 
composed of weakly coupled layers. It is only recently that the synthesis of two-dimensional magnetic compounds, 
such as those based on rare-earth oxide YbMgGaO$_4$\cite{prx10.011007,natcommun10.4530}, has made it possible
to observe the occurrence of the BKT transition in magnetic models\cite{natcommun11.1}. 

Contrary to the easy-plane scenario, we have the 2D magnetic models with easy-axis anisotropy, which
break the continuous internal symmetry and provide an Ising-type transition. It should be noted that in this case, 
the Mermin-Wagner theorem does not apply, and even a minor anisotropy is sufficient to induce a finite transition
temperature \cite{prb38.12015,prb59.6229,prb62.3771}. In addition, since the first synthesis of graphene
monolayers\cite{scince306.5696}, there has been tremendous interest in utilizing low-dimensional materials 
for technological applications. Specifically, numerous studies focused on the deployment of spintronic devices, 
which employ spin rather than electronic charge for processing and storing information\cite{npj2d4.17,natnano15.545}. 
However, graphene is a strongly diamagnetic material and shows a weak spin-orbit coupling\cite{rmp92.021003}, which hinders 
its effectiveness as a spin current detector via Inverse Spin-Hall Effect\cite{apl88.182509}. On the other hand, the recent 
development of the 2D van der Waals magnets\cite{nature546.265,nature563.47,apl8.031305} has further intensified 
research in low-dimensional magnets. For instance, Torelli et al. used Monte Carlo simulation models to 
investigate the role of anisotropy in several two-dimensional magnets\cite{2dmater4.045018}, some of which exhibit
order even at room temperature.

As it is well-known, there is no theoretical method capable of yielding exact results in two-dimensional magnetic 
models, and approximations are required to extract any physical information about the system. Among the most
traditional methods, we have bosonic representations such as the Holstein-Primakoff (HP), Schwinger, and Dyson-Maleev
formalism, which replace the spin by magnon annihilation/creation operators (for a review of bosonic representations,
and other methods, see Ref. \cite{auerbach,pires}. Each method is more or less 
convenient depending on the magnetic system properties, but in the simpler description of non-interacting spin waves, 
they all result in quadratic models. However, eventually, the traditional quadratic representation is not suficient
to describe the properties even as an initial approximation. In the spin conductivity evaluation, for example, the
first non-null contributions arise from terms associated with interacting magnons\cite{prb75.214403}.
These contributions necessitate the inclusion of terms beyond the second order in the Hamiltonian. 
In these scenarios, an extensive perturbative analysis is required, and the solution can be hard to be obtained. 
Therefore, the SCHA formalism emerges as an alternative to the traditional bosonic representations in order 
to address these complexities.

The main idea of the SCHA is the replacement of the original spin Hamiltonian by another one containing only quadratic 
terms, similar to the aforementioned methods, involving the canonically conjugate fields (operators, in 
the quantum approach) $\varphi$ and $S^z$. However, unlike the Holstein-Primakoff formalism, the SCHA includes
spin fluctuations through self-consistently solved renormalization parameters. Therefore, the SCHA model is simultaneously
simple and precise in determining the thermodynamics of ordered phases in magnetic materials. Indeed, over the years, the SCHA 
has been successfully used to determine the critical temperature
\cite{prb49.9663,pla202.309,prb51.16413,prb54.3019,ssc104.771,prb59.6229}, the topological BKT transition \cite{pla166.330,prb48.12698,prb49.9663,prb50.9592,ssc100.791,prb53.235,prb54.3019,prb54.6081,ssc112.705,epjb2.169,pssb.242.2138,prb78.212408,jmmm452.315}, and the large-D quantum phase 
transition \cite{pasma373.387,jpcm20.015208,pasma388.21,pasma388.3779,jmmm357.45} in a wide variety of magnetic models.
In addition, the description by using canonically conjugate fields supply the most convenient formalism to 
describe coherent states in magnetism, as demonstrated by Moura and Lopes\cite{jmmm472.1}. More recently, Moura has 
applied the same formalism to present a detailed theoretical analysis of the ferromagnetic resonance and spin pumping 
process at the ferromagnetic/normal metal junction\cite{prb106.054313}. 

In this paper, we employ the SCHA to analyze ferro and antiferromagnetic 2D models exhibiting easy-axis anisotropy. 
While the method has been extensively utilized to investigate magnetic models with easy-plane anisotropy, we have a 
dearth of studies focusing on the effects of easy-axis anisotropy. Therefore, we developed the SCHA formalism for determining 
the properties of anisotropic two-dimensional magnetic models, which are given by the Hamiltonian
\begin{equation}
\label{eq.Hamiltonian}
H=\pm J\sum_{\langle i,j\rangle}(S_i^y S_j^y+S_i^z S_j^z+\lambda_0 S_i^x S_j^x)-g\mu_B\sum_i {\bf S_i}\cdot{\bf B_i},
\end{equation}
where $-J$ ($+J$) represents the ferromagnetic (antiferromagnetic) coupling, $\lambda_0$ is the (bare) anisotropic constant, 
and the first sum is carried out over nearest-neighbor sites. 
The last term is the Zeeman energy associated with the magnetic field ${\bf B}$. For a thin film, 
we can write ${\bf B}=\mu_0 ({\bf H_\textrm{ext}}-N_x{\bf M_x})$, where ${\bf H_\textrm{ext}}$ is the
external $H$-field and $-N_x{\bf M_x}$ is the demagnetizing field. Here, we are considering the special case with uniform magnetization
along the x-axis that provides the demagnetization factors $N_x=1$, while $N_y=N_z=0$. In the antiferromagnetic case,
${\bf M}={\bf 0}$, and there is no demagnetizing field to be considered. For ferromagnetic models, the demagnetizing field effects
are considered replacing $\lambda_0$ by an effective anisotropy, as demonstrated in the next section.
To facilitate the application of the SCHA formalism, we consider sites located on the yz-plane while the anisotropy is 
defined along the x-axis. The easy-axis anisotropy is achieved by setting $\lambda>1$, and the limit $\lambda$ approaching infinity 
corresponds to the Ising model. Alternatively, a uniaxial anisotropy, which also represents easy-axis anisotropic systems, 
could be considered in replacement of $\lambda S_i^x S_j^x$ with minor modifications. 
Moreover, other interactions or anisotropies can be easily implemented, enabling the application
of formalism in a wide variety of situations. The SCHA proves to be a valuable tool for determining essential properties 
of the ordered phase, such as the spectrum energy, the transition temperature, the magnetization, and others. The results 
obtained from SCHA exhibit excellent agreement with numerous experimental and Monte Carlo simulation studies. 

\section{The SCHA formalism}
\begin{figure}[h]
\centering \epsfig{file=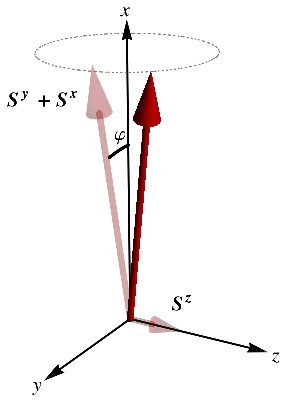,width=0.5\linewidth}
\caption{The quantization axis is adopted along the $x$ axis. $S^z$ and $\varphi$ are canonically conjugate, and
$\varphi$ is defined as the angle between the spin projection on the $xy$ plane and the $x$ axis.}
\label{fig.single_spin}
\end{figure}

As mentioned earlier, the initial step of the SCHA formalism is replacing the spin fields with quadratic representations
of angle $\varphi$ around the z-axis and its conjugate momentum, denoted as $S^z$. Typically, in problems covering planar 
anisotropy, we consider spins located on the xy-plane and define the quantization along the z-axis. However, this choice
is not mandatory. In the present context, to properly apply the SCHA formalism, we opt to place the spins on the yz-plane, 
while retaining the x-axis as the magnetization axis, as shown in Fig. (\ref{fig.single_spin}). 
Due to the anisotropy, $\varphi\ll 1$ and thus, the transverse spin components $S^y$ and $S^z$ are much smaller 
than the longitudinal component $S^x$. From the classical point of view, the fields $\varphi_i$ and $S_j^z$, 
on sites i and j, respectively, satisfy the Poisson bracket $\{\varphi_i,S_j^z\}=\delta_{ij}$, and 
the quantization is achieved by promoting the fields to operators that obey the commutation relation 
$[\varphi_i,S_j^z]=\delta_{ij}$.

We can consider the spin as a quantum operator from the beginning, and then use the Villain representation\cite{jp35.27} to 
represent the raising operator as $S_i^+=e^{i\varphi}\sqrt{\tilde{S}^2-S_i^z(S_i^z+1)}$, where $\tilde{S}=\sqrt{S(S+1)}$. 
Alternatively, we can start in the semiclassical limit, for which the spins are represented by three-dimensional
vectors. Then, we perform the quantization by the usual second quantization procedure. Since the second approach is 
more convenient for the present scenario, we will proceed with this approach.

\subsection*{Semiclassical approach}
To begin with, we write the Hamiltonian (\ref{eq.Hamiltonian}) in terms of the $\varphi$ and $S^z$ fields, which results in
\begin{IEEEeqnarray}{rCl}
\label{eq.H}
H&=&\frac{J}{2}\sum_{\langle ij\rangle}\left[-(1+\lambda_0)f_{ij}\cos\Delta\varphi+(1-\lambda_0)f_{ij}\cos\Sigma\varphi\pm\right.\nonumber\\
&&\left.\pm 2S_i^z S_j^z\right]-g\mu_B B^x\sum_i \sqrt{f_{ii}}\cos\varphi_i,
\end{IEEEeqnarray}
where $f_{ij}=\sqrt{[S^2-(S_i^z)^2][S^2-(S_j^z)^2]}$, $\Delta\varphi=\varphi_j-\varphi_i$, and $\Sigma\varphi=\varphi_j+\varphi_i$. 
In the above equation, for the antiferromagnetic model, we apply a rotation of $\pi$ around the z-axis for one of the sublattices. 
The quadratic Hamiltonian $H_0$ is then obtained by expanding the cosines and the $f{ij}$ function in terms of $\varphi$ and $S^z$.
The dynamics of $S^z$ relies on the choice of Hamiltonian used to determine the time derivative, whether it is $H$ or $H_0$.
The resulting dynamics are strongly affected by whether the quadratic Hamiltonian or the complicated full version is used. 
To address this issue, we introduce renormalization parameters 
$\rho_\Delta$, $\rho_\Sigma$, and $\rho_x$ replacing the cosine functions 
$\cos\Delta\varphi$, $\cos\Sigma\varphi$, and $\cos\varphi_i$ with the terms $-\rho_\Delta\Delta\varphi^2/2$, $-\rho_\Sigma\Sigma\varphi^2/2$, 
and $-\rho_x \varphi^2/2$, respectively. The determination of the renormalization parameters aims to optimize 
the harmonic model, providing the most accurate representation of the system's behavior. 
Therefore, the quadratic Hamiltonian, in momentum space, is expressed as
\begin{equation}
\label{eq.H0}
H_0=\frac{1}{2}\sum_q\left(h_q^\varphi \bar{\varphi}_q\varphi_q+h_q^z\bar{S}_q^z S_q^z\right),
\end{equation}
where $\bar{\varphi}_q=\varphi_{-q}$, $\bar{S}_q^z=S_{-q}^z$. The matrix coefficients are expressed as
$h_q^\varphi=zJS^2(\lambda\rho_x-\rho_y\gamma_q)$ and $h_q^z=zJ(\lambda\pm\gamma_q)$. Here, $\lambda=\lambda_0+\lambda_M$,
with $\lambda_M=g\mu_B\mu_0M_x/S$. Note that, for antiferromagnetic models, $M_x=0$ and $\lambda=\lambda_0$.
The renormalization parameters $\rho_{x,y}$ are based on an alternative expansion of 
the angular part. By adopting the ansatz $S^y\approx S\sqrt{\rho_y}\varphi$ and $S^x\approx S-(S^z)^2/2S-S\rho_x\varphi^2/2$, 
it is a straightforward procedure to demonstrate that the revised expansion yields the same Hamiltonian $H_0$ with 
\begin{IEEEeqnarray}{rCl}
\IEEEyesnumber
\IEEEyessubnumber*
\rho_x=\frac{(\lambda+1)\rho_\Delta+(\lambda-1)\rho_\Sigma}{2\lambda},\\
\rho_y=\frac{(\lambda+1)\rho_\Delta-(\lambda-1)\rho_\Sigma}{2}.
\end{IEEEeqnarray}
For the square lattice, the structure factor is written as $\gamma_q=\cos(q_x/2)\cos(q_y/2)$, considering the
unitary lattice parameter, and the number of nearest neighbors is $z=4$. 
In Appendix \ref{appendix}, we show that the renormalization parameters are obtained from two coupled
self-consistent equations given by
\begin{IEEEeqnarray}{rCl}
\label{eq.rho}
\IEEEyesnumber
\IEEEyessubnumber*
\rho_\Delta&=&\left(1-\frac{\langle(S^z)^2\rangle_0}{S^2}\right)\exp\left(-\frac{\langle\Delta\varphi^2\rangle_0}{2}\right)\\
\rho_\Sigma&=&\left(1-\frac{\langle(S^z)^2\rangle_0}{S^2}\right)\exp\left(-\frac{\langle\Sigma\varphi^2\rangle_0}{2}\right),
\end{IEEEeqnarray}
where we use the harmonic Hamiltonian $H_0$ for evaluating the averages. In the semiclassical limit, the Hamiltonian consists 
of two independent fields, and the thermodynamic averages can be easily determined. Extending the integration limit
to $-\infty<\varphi,S^z<\infty$, we deal with Gaussian integrals that provide
\begin{equation}
\label{eq.SzSz}
\langle(S^z)^2\rangle_0=S^2\int_\textrm{BZ} \frac{d^2q}{A_\textrm{BZ}} \frac{t}{z(\lambda\pm\gamma_q)},
\end{equation} 
\begin{equation}
\label{eq.Deltaphi2}
\langle(\Delta\varphi)^2\rangle_0=\int_\textrm{BZ} \frac{d^2q}{A_\textrm{BZ}} \frac{2(1-\gamma_q)t}{z(\lambda\rho_x-\rho_y\gamma_q)},
\end{equation}
and
\begin{equation}
\label{eq.Sigmaphi2}
\langle(\Sigma\varphi)^2\rangle_0=\int_\textrm{BZ} \frac{d^2q}{A_\textrm{BZ}} \frac{2(1+\gamma_q)t}{z(\lambda\rho_x-\rho_y\gamma_q)},
\end{equation}
where we define the dimensionless reduced temperature $t=k_B T/JS^2$. 
The integrals are carried out in the first Brillouin zone with the correspondent area denoted as $A_\textrm{BZ}=8\pi^2$.

\begin{figure}[h]
\centering \epsfig{file=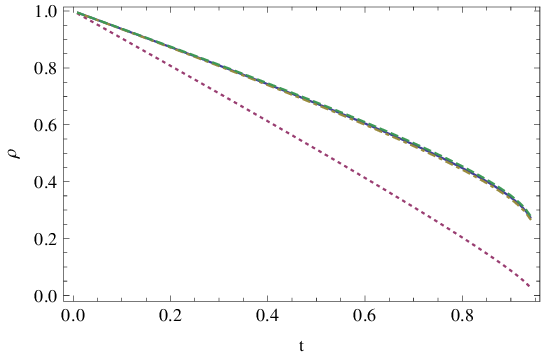,width=1.0\linewidth}
\caption{Temperature dependence of the renormalization parameters $\rho_\Delta$, $\rho_\Sigma$, $\rho_x$, 
and $\rho_y$, for a classical antiferromagnetic model with $\lambda=1.05$. The self-consistent equations yield the reduced 
critical temperature $t_c=0.94$. The (purple) dotted line represents $\rho_\Sigma$, while $\rho_x$, $\rho_y$, and $\rho_\Delta$
are given approximately by the same line.}
\label{fig.rho_classical}
\end{figure}

Note that, for small anisotropies, $\delta\lambda=\lambda-1\ll 1$, the deviation between the parameters 
$\rho_\Delta$, $\rho_\Sigma$, and $\rho_x$ is negligible. Indeed, for small $\delta\lambda$, we
can write $\rho_x=\rho_\Delta-\delta\lambda(\rho_\Delta-\rho_\Sigma)/2$ and $\rho_y=\rho_\Delta+\delta\lambda(\rho_\Delta-\rho_\Sigma)/2$.
Fig. (\ref{fig.rho_classical}) shows the dependence on the temperature for the renormalization parameters 
$\rho_\Delta$, $\rho_\Sigma$, $\rho_x$, and $\rho_y$, in an antiferromagnetic model with $\lambda=1.05$. 
For this case, the critical temperature provided by the self-consistent equations is $t_c=0.94$. Using Eq. (\ref{eq.tc}), 
demonstrated below, the critical temperature is given by $t_c=0.97$.

The spin dynamics is obtained from the Hamilton equations $\hbar\dot\varphi_q=-\partial H_0/\partial\bar{S}_q^z$ and
$\hbar\dot{S}_q^z=\partial H_0/\partial \bar{\varphi}_q$, which yield the coupled transverse spin equations
\begin{IEEEeqnarray}{rCl}
\IEEEyesnumber
\IEEEyessubnumber*
\hbar\dot{S}_q^y&=&-(S\sqrt{\rho_y})h_q^z S_q^z,\\
\hbar\dot{S}_q^z&=&(S\sqrt{\rho_y})^{-1}h_q^\varphi S_q^y.
\end{IEEEeqnarray}
By adopting the typical elliptical oscillating transverse behavior of the spin field, characterized by the solutions
$S_q^y(t)=A_q^y\cos(\omega_q t)$ and $S_q^z(t)=A_q^z\sin(\omega_q t)$, with $A_q^{y,z}$ representing the
the transverse amplitudes, we obtain the spin-wave energy $\epsilon_q=\hbar\omega_q=\sqrt{h_q^\varphi h_q^z}$. At 
the low-temperature limit, the renormalization has a minor effect, and we can assume $\rho_x\approx\rho_y=\rho$ (at
$T=0$, it is easy to show that $\rho=1$). 
Therefore, the spin-wave energy takes on a simpler form denoted as $\epsilon_q=zJ_rS(\lambda-\gamma_q)$ for ferromagnetic 
insulator, and $\epsilon_q=zJ_r S\sqrt{\lambda^2-\gamma_q^2}$ for antiferromagnetic one, where we define the temperature-dependent
renormalized coupling $J_r(T)=\sqrt{\rho(T)}J$. Note that $J_r$ represents the exchange coupling obtained from
experiments at finite temperatures, while the bare value $J$ is the physical parameter only at zero temperature.
In addition, due to the anisotropy, both cases show an energy gap, 
which vanishes when $\lambda=1$, that suppresses the spin fluctuations resulting in a finite transition temperature. In the 
long-wavelength limit, the ferromagnetic model shows the usual non-relativistic behavior with 
$\epsilon_q\approx 4J_rS\delta\lambda+J_rSq^2$, where $\delta\lambda=\lambda-1$ represents the deviation from the isotropic limit. 
Considering $\delta\lambda\ll 1$, the antiferromagnet also presents the non-relativistic energy
$\epsilon_q=4J_rS\sqrt{2\delta\lambda}+(J_rS/\sqrt{2\delta\lambda})q^2$ in contrast to the relativistic spectrum 
$\epsilon_q=(2\sqrt{2}J_rS)q$ obtained in the isotropic model.

At finite temperatures, the evaluation of the renormalization parameters entails the resolution of the system of coupled equations 
given by Eq. (\ref{eq.rho}). Generally, we perform this assessment through the utilization of self-consistently numeric integration
techniques. Nevertheless, under the condition of small anisotropy ($\delta\lambda\ll1$), we can acquire an analytical 
expression for $t_c$. In this case, $\rho_x\approx\rho_y=\rho$, even at high temperatures, close to $T_c$. Therefore, we can simplify 
Eq. ({\ref{eq.Deltaphi2}) for obtaining $\langle\Delta\phi^2\rangle_0=[1-\delta\lambda I(\delta\lambda)]t/2\rho$, where
\begin{equation}
\label{eq.I}
I(\delta\lambda)=\int_\textrm{BZ}\frac{d^2q}{A_\textrm{BZ}}\frac{1}{1+\delta\lambda\pm\gamma_q}
\end{equation}
The self-consistent equation exhibits an abrupt non-physical vanishing at the critical temperature $t_c$. 
The same behavior is observed in the magnetization obtained from SCHA, as described subsequently. Consequently, this phenomenon 
leads to an erroneous first-order transition for magnetization. Such anomalies are frequently encountered in theories 
based on harmonic expansions. Nevertheless, despite this limitation, for temperatures $t<t_c$, the SCHA method provides 
excellent results, while the formalism exhibits an issue only at temperatures very close to $tc$. It is worth noting that even for 
traditional spin representations, such as the Holstein-Primakoff formalism, it is difficult to properly describe the thermodynamics close 
to the critical temperature. Despite the excellent agreement in the low-temperature regime, when adopting the non-interacting spin-wave 
limit, the HP representation provides poor results for $T_c$. In this scenario, it is necessary to incorporate the quartic-order 
terms, which renormalize the spin-wave energy and thereby furnish a more reasonable critical temperature\cite{pps82.992}.
On the other hand, the critical temperature $t_c$ obtained from the harmonic approximation is very close to the real
critical temperature. Therefore, notwithstanding the imprecise behavior at $t=t_c$, we will consider $t_c$ as a 
reasonable estimation for the critical temperature. Using the condition $dt/d\rho=0$, at the
critical temperature, it is a straightforward procedure to reach the result
\begin{equation}
\label{eq.tc}
t_c(\delta\lambda)=\frac{4}{e+(1-e \delta\lambda)I(\delta\lambda)},
\end{equation}
where $e$ is the base of the natural logarithm. In Sec. (\ref{sec.app}), we will apply Eq. (\ref{eq.tc}) to determine the 
critical temperature of many experimental and Monte Carlo simulation studies.

Due to the easy-axis anisotropy, the spontaneous magnetization occurs in the direction perpendicular to the yz-plane and
$M=\langle S^x\rangle$. Taking into account the uncoupled harmonic fields $\varphi$ and $S^z$,
the average $\langle S^x\rangle$ for the ferromagnetic model is simply expressed as
\begin{equation}
\frac{M_\textrm{FM}}{M_s}=\left(1-\frac{\langle(S^z)^2\rangle_0}{2S^2}\right)\exp\left(-\frac{\langle\varphi^2\rangle_0}{2}\right),
\end{equation} 
where we expand the square root up to first-order and use the equality $\langle\cos\varphi\rangle_0=\exp(-\langle\varphi^2\rangle_0/2)$,
valid for Gaussian distributions. In the above equation, $M_s=S$ represents the saturation magnetization. For the antiferromagnetic model, 
it is necessary to analyze each sublattice separately, which results in
\begin{equation}
\frac{M_\textrm{AFM}}{M_s}=\left(1-\frac{\langle(S^z)^2\rangle_0}{4S^2}\right)\exp\left(-\frac{\langle\varphi^2\rangle_0}{4}\right),
\end{equation} 
where the averages are reduced by two because the sublattice average takes into account only half of the sites.

\subsection*{Quantum approach}
Although the semiclassical approach proves highly effective in solving large spin models, it is imperative to treat 
models featuring half-integer spin as purely quantum models.  In the case of planar anisotropy, the quantum SCHA has 
been formulated by considering the spins in Hamiltonian either as operators\cite{jp35.27,ssc104.771} or 
through the quantization of classical fields\cite{prb48.12698}. Regardless of which mode we used, the final 
results coincide. Hence, a resembling construction is expected for the easy-axis anisotropic model, and given
our prior development of the classical SCHA, we proceed to quantize the Hamiltonian (\ref{eq.H0}) following the 
conventional procedure of second quantization. 

Within the quantum framework, the fields $\varphi$ and $S^z$ are replaced by operators that satisfy the commutation
relation $[\varphi_i,S_j^z]=\delta_{ij}$. Consequently, it becomes advantageous to represent the Hamiltonian in terms 
of annihilation and creation operators. In favor of the bosonic operators $a_q$ and $a_q^\dagger$, we perform the 
transformation
\begin{IEEEeqnarray}{rCl}
\IEEEyesnumber
\IEEEyessubnumber*
\varphi_q&=&\frac{1}{\sqrt{2}}\left(\frac{h_q^z}{h_q^\varphi}\right)^{1/4}(a_q^\dagger+a_{-q}),\\
S_q^z&=&\frac{i}{\sqrt{2}}\left(\frac{h_q^\varphi}{h_q^z}\right)^{1/4}(a_q^\dagger-a_{-q}),
\end{IEEEeqnarray}
to yield the quantum harmonic Hamiltonian $H_0=\sum_q \epsilon_q (a_q^\dagger a_q+1/2)$, where the magnon energy
is $\epsilon_q=\sqrt{h_q^\varphi h_q^z}$ (identical to the spin-wave energy obtained from the semiclassical analysis).
The self-consistent equations are the same as given by Eq. (\ref{eq.rho}) except that $S$ is replaced by $\tilde{S}=\sqrt{S(S+1)}$, 
and the averages must be determined by using the magnon number states $|n_q\rangle$ (eigenvalues of $H_0$). Therefore, we obtain
\begin{equation}
\langle S_{-q}^zS_q^z\rangle_0=\frac{1}{2}\sqrt{\frac{h_q^\varphi}{h_q^z}}\coth\left(\frac{\beta\epsilon_q}{2}\right),
\end{equation}
and
\begin{equation}
\langle \varphi_{-q}\varphi_q\rangle_0=\frac{1}{2}\sqrt{\frac{h_q^z}{h_q^\varphi}}\coth\left(\frac{\beta\epsilon_q}{2}\right).
\end{equation}
Note that, in the semiclassical limit, $\beta\epsilon_q\ll 1$, and the series expansion $\coth(\beta\epsilon_q/2)\approx 2/\beta\epsilon_q$
provides the same averages as those obtained in the previous section. However, due to the complexity of the average equations, it is not
feasible to derive a quantum version of Eq. ({\ref{eq.tc}). Therefore, we can determine the critical temperature only by 
solving the quantum self-consistent equations. Furthermore, quantum fluctuations persist even at zero temperature, resulting in 
significant reductions in the renormalization parameters across the entire temperature range, as shown in Fig. ({\ref{fig.rho_quantum}).

\begin{figure}[h]
\centering \epsfig{file=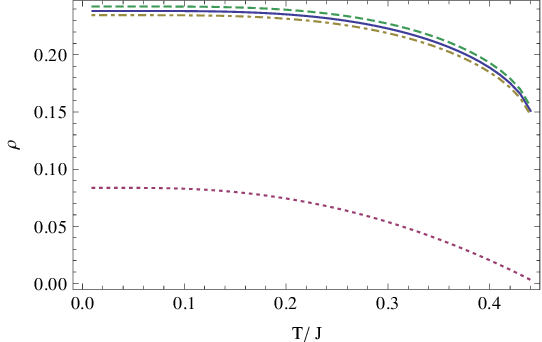,width=1.0\linewidth}
\caption{Temperature dependence of the renormalization parameters $\rho_\Delta$, $\rho_\Sigma$, $\rho_x$, 
and $\rho_y$, for a quantum antiferromagnetic model with $\lambda=1.05$ and $S=1/2$.
The (purple) dotted line represents $\rho_\Sigma$, while $\rho_x$, $\rho_y$, and $\rho_\Delta$
are given approximately by the same line. Here, the critical temperature is given by $T_c=0.44J$.}
\label{fig.rho_quantum}
\end{figure}

\section{Applications of the SCHA}
\label{sec.app}
In this section, we employ the SCHA for describing a large number of studies involving easy-axis anisotropic models.
In the description of two-dimensional ferromagnetic materials, certain experimental investigations have 
adopted a theory based on an external staggered magnetic field, which breaks
the symmetry and provides ordering across all temperatures, as a means to describe the anisotropic behavior.
However, following Cuccoli {\it et. al.}\cite{prb62.3771}, we favor the utilization of the easy-axis 
anisotropy, as it aligns more closely with physical plausibility and yields satisfactory outcomes. 

In the subsequent paragraphs, we utilize the SCHA to analyze some experimental data of antiferromagnetic systems. Among these
systems, Rb$_2$MnF$_4$\cite{prb1.2211,jap42.1595,prb8.285,prb15.4292}, Rb$_2$MnCl$_4$\cite{pssb97.501,jpcssp19.4503}, 
K$_2$MnF$_4$\cite{prb8.285,prb8.304}, and (CH$_3$NH$_3$)$_2$MnCl$_4$\cite{pssb97.501}
exhibit a square crystalline lattice and possess large spin value of $S=5/2$, which allow the effective application of 
the semiclassical approach. In addition, we include an $S=1$ model given by the K$_2$NiF$_4$]\cite{prl23.1394,prb3.1736,prb8.285} 
compound. For the $S=1$ spin value, the semiclassical approach does not provide good results as occurs for the $S=5/2$ case,
and a quantum analysis is necessary. 

For a proper description, instead of using a pre-determined coupling constant and
anisotropy, which are determined for other models rather than SCHA, we use the spectrum energy data to find the best
parameters to fit the experimental measurement. For antiferromagnetic models, the gap is expressed as 
$\Delta=zJ_r S\sqrt{\lambda^2-1}$, which provides the anisotropy deviation $\delta\lambda=(1/2)(\Delta/zJ_r S)^2$. 
The renormalized coupling, at temperature $T_0$, is then 
determined by applying the minimum square method to the error $\sum_q(\epsilon_q-\epsilon_q^\textrm{exp})^2$, 
where $\epsilon_q^\textrm{exp}$ is the experimental energy data, measured at $T_0$, and the momentum sum 
is done over the experimental values. The minimization in relation to $J_r$ yields
\begin{equation}
J_r=\frac{1}{zS}\frac{\sum_q \epsilon_q^\textrm{exp}\sqrt{1+2\delta\lambda-\gamma_q^2}}{\sum_q(1+2\delta\lambda-\gamma_q^2)},
\end{equation}
which is self-consistently solved to find $J_r(T_0)$. Finally, we solve Eq. (\ref{eq.rho}) at $T_0$ to determine
the bare coupling constant $J=J_r/\sqrt{\rho}$, which is temperature independent and used to obtain the SCHA outcomes
that are listed in Table (\ref{table.tc}). Since $\delta\lambda\ll 1$, the critical temperature is evaluated through 
the self-consistent equations and using Eq. ({\ref{eq.tc}). Details about each one of the compounds are listed below.

{\bf Rb$_2$MnF$_4$} - Based on Ref. \cite{prb15.4292}, we obtain the gap energy $\Delta(T_0)=0.510$ meV (5.92 K), 
measured at $T_0=4.2$ K, which provides $\lambda=1.00284$. The bare coupling constant calculated is given by $J=8.14$ K. 
The critical temperature obtained from SCHA is in remarkable agreement with the experimental data.

{\bf K$_2$MnF$_4$} - From the Ref. {\cite{prb8.304}, we extract the energy gap $\Delta(T_0)=0.649$ meV (7.53 K)
at $T_0=4.5$ K. The anisotropy is determined as $\lambda=1.00372$, while the exchange constante is $J=9.05$ K. In this 
case, the evaluated critical temperature is also in good agreement with the real value.

{\bf Rb$_2$MnCl$_4$} - The data from Ref. \cite{pssb97.501}, gives the energy gap $\Delta(T_0)=11.38$ K, 
determined at $T_0=8$ K. For $S=5/2$, we find $\lambda=1.00466$ and $J=11.80$ K, which results in a critical temperature
of approximately 9 percent larger than the experimental value. 

{\bf MAMC} - The organic chloride compound (CH$_3$NH$_3$)$_2$MnCl$_4$ (abbreviated MAMC) shows a structure similar
to the Rb$_2$MnCl$_4$. Using the data of Ref. \cite{pssb97.501}, measured at $T_0=8.5$ K, we obtain $J=10.28$ K and 
$\lambda=1.00336$, using $S=5/2$. In this case, the critical temperature obtained from SCHA is almost 10 percent larger than
the real value.

{\bf K$_2$NiF$_4$} - This is the one analyzed compound with $S=1$. Due to the small spin value, the semiclassical provides
only a qualitative result, and better results are achieved through the quantum approach. 
From the Ref. (\cite{prb3.1736}), we get the energy gap $\Delta=27.70$ meV, measured at $T_0=5$ K, which results in
$J=114.49$ K and in $\lambda=1.00175$. 

\begin{table}[h]
\centering
\begin{tabular}{l  l  l  l  l  l}
\hline\hline
Compound				& Spin  & Eq. (\ref{eq.tc}) & SC-SCHA & Q-SCHA & Exp. \\
\hline
Rb$_2$MnF$_4$		& 5/2   & 39.0 K & 38.7 K & 36.4 K & 38.4 K$^a$\\
K$_2$MnF$_4$	 		& 5/2   & 44.1 K & 43.6 K & 41.0 K & 42.1 K$^b$\\
MAMC 				& 5/2   & 49.7 K & 49.5 K & 45.3 K & 45.3 K$^c$\\
					& 2.28  & 45.6 K & 45.4 K & 42.1 K & 45.3 K\\
Rb$_2$MnCl$_4$ 		& 5/2   & 61.1 K & 60.2 K & 56.0 K & 56.0 K$^c$\\
					& 2.32  & 56.9 K & 56.1 K & 52.8 K & 56.0 K\\
K$_2$NiF$_4$			& 1	    & 85.7 K & 84.7 K & 98.1 K & 97.2 K$^d$		\\					
\hline
\end{tabular}
\label{table.tc}
\caption{The results of the SCHA analysis are presented herein. The critical temperature was determined by applying Eq. (\ref{eq.tc}),
using the numeric results of the semiclassical approach (SC-SCHA) as well as the quantum SCHA formalism (Q-SCHA). 
Additionally, for Rb$_2$MnCl$_4$ and MAMC, we display the critical temperature calculated by considering the spin value 
$S = 5/2$, along with an effective value estimated using Eq. (\ref{eq.tc}). \\ $^a$ Phys. Rev. B 15, 4292 (1977)\\
$^b$ Phys. Rev. B 8, 304 (1973)\\ $^c$ Phys. Status Solidi (B) 97, 501 (1980)\\ $^d$ Phys. Rev. B 3, 1736 (1971)}
\end{table}

\begin{figure}[h]
\centering \epsfig{file=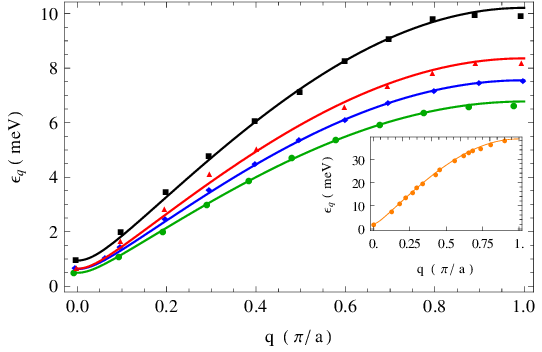,width=\linewidth}
\caption{Spin-wave energy for Rb$_2$MnF$_4$ (circles in green), K$_2$MnF$_4$ (diamonds in blue), MAMC (triangles in red), 
and Rb$_2$MnCl$_4$ (squares in black), measured along the ${\bf q}=(q,0,0)$ direction. Solid lines represent the SCHA results, 
while points are experimental data from the references cited in the main text. Here, the solid lines are determined by 
adopting $S=5/2$ for all compounds. Additionally, the inset plot presents the energy spectrum for K$_2$NiF$_4$.} 
\label{fig.spectrum}
\end{figure}

Fig. (\ref{fig.spectrum}) shows the energy spectrum for the four compounds listed in Table (\ref{table.tc}),
taking into account a spin value of $S=5/2$. The parameters $J$ and $\lambda$ were adjusted to achieve the optimal fit 
between theoretical predictions and experimental data. The SCHA formalism yields a notable agreement, as observed.

The critical temperature $T_c$ for different compounds, namely Rb$_2$MnF$_4$, K$_2$MnF$_4$, Rb$_2$MnCl$_4$, and MAMC, 
was determined using the SCHA. The results indicate that $T_c$ is in notable agreement with experimental values for 
Rb$_2$MnF$_4$ and K$_2$MnF$_4$, while for Rb$_2$MnCl$_4$ and MAMC, the calculated $T_c$ is approximately 10 percent higher 
than the actual value.  Generally, theoretical methods are better for describing Rb$_2$MnF$_4$ and K$_2$MnF$_4$. 
For instance, Pich and Schwabl applied the method of Callen to evaluate 
Green's function for the same compounds listed in Table (\ref{table.tc}}) and they also obtained better agreement 
for Rb$_2$MnF$_4$ and K$_2$MnF$_4$ rather than Rb$_2$MnCl$_4$ and MAMC\cite{prb49.413,jmmm140.1709}. In their studies, the spin $S$
was replaced by an effective spin $S_\textrm{eff}=2.32$, accounting for quantum and thermal renormalizations.
Cuccoli {\it et. al.} have applied the pure-quantum self-consistent harmonic approximation \cite{pra45.8418}, a different harmonic 
approximation based on Weyl quantization for determining thermodynamic averages, to successfully estimate the critical temperature of 
Rb$_2$MnF$_4$\cite{prb62.3771}. Conversely, Random Phase Approximation formalism have been used for Rb$_2$MnCl$_4$\cite{epjb68.511} and 
MAMC\cite{ssc151.1753} to obtain more precise values for $T_c$. 

To improve the results for Rb$_2$MnCl$_4$ and MAMC, we also adopt an effective spin $S_\textrm{eff}$ in place of $S$, 
as explored in Ref. \cite{prb49.413}. Taking an interesting perspective, K\"{o}bler has considered the interaction between 
magnons and Glashow-Weinberg-Salam (GWS) bosons of the magnetic continuum field to explain the energy gap and 
the behavior in the long-wavelength limit of various magnetic models\cite{appa127.1694}. According to this approach, due to 
the magnon-GWS boson interaction, the spin of Rb$_2$MnCl$_4$ was reduced from $S=5/2$ to $S_\textrm{eff}=2$. Here, 
we used Eq. ({\ref{eq.tc}) to estimate the effective spin that results in a critical temperature close to the experimental values, 
obtaining $S_\textrm{eff}=2.28$ for MAMC, and $S_\textrm{eff}=2.32$ for Rb$_2$MnCl$_4$, close to the values of Ref.\cite{prb49.413}. 
Both critical temperature values, using $S=5/2$ and $S_\textrm{eff}$ are listed in Table \ref{table.tc}.

\begin{figure}[h]
\centering \epsfig{file=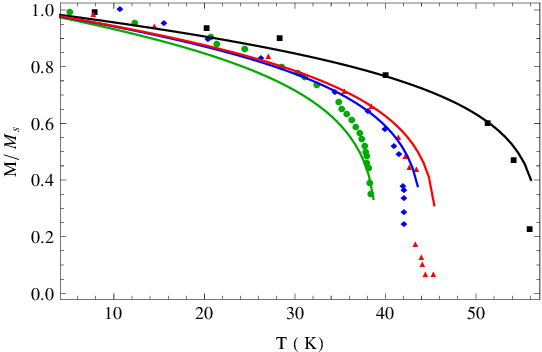,width=\linewidth}
\caption{Dependence on temperature of the magnetization for Rb$_2$MnF$_4$ (circles in green), K$_2$MnF$_4$ (diamonds in blue), 
MAMC (triangles in red), and Rb$_2$MnCl$_4$ (squares in black). Solid lines represent the SCHA results while points are experimental data from
the references cited in main text. Here, Rb$_2$MnF$_4$ and K$_2$MnF$_4$ outcomes are determined using $S=5/2$, while for Rb$_2$MnCl$_4$
and MAMC we adopt effective spin values for critical temperature with better experimental agreement.} 
\label{fig.magnetization}
\end{figure}

Fig. (\ref{fig.magnetization}) shows the magnetization dependence on temperature for the four $S=5/2$ compounds. The 
outcomes derived from the SCHA exhibit a moderate level of concordance with experimental data. 
Specifically, when applied to the compound Rb$_2$MnF$_4$, the agreement between theoretical predictions and experimental 
observations is notably less pronounced. A potential remedy for this discrepancy is given by incorporating spin interactions 
that extend beyond the nearest-neighbors coupling. Cowley \textit{et al.} adopted interactions between nearest and 
next-nearest neighbors to describe Rb$_2$MnF$_4$\cite{prb15.4292}. This extended model yielded agreement with experimental results that 
was threefold superior to the conventional model, which solely considers nearest-neighbor coupling. 	
It is noteworthy, however, that including these additional interactions within the SCHA formalism falls outside 
the scope of the present study. Considering the power-law behavior for the 
magnetization $M(T)/M_s\propto (1-T/T_c)^\beta$ over the entire temperature interval, we obtain an almost uniform $\beta$ 
parameter, varying between $\beta=0.19$ (Rb$_2$MnCl$_4$) up to $\beta=0.23$ (K$_2$NiF$_4$). It is well-known, from the
critical phenomena theory, that the power-law behavior is expected only close to the critical temperature, and the $\beta$
parameter must be determined around $T_c$. Adopting the interval $0.9T_c\leq T< T_c$, we use the SCHA results for
determining the critical exponent, and the obtained values are exhibited in Table (\ref{table.beta}). 
Provided the complicated behavior close to the critical temperature, it is not expectable
a precise result for $\beta$ and so, the critical exponents show only a moderate agreement with experiments. The same
problem is also obtained by using other theoretical formalisms, and a precise analysis at the critical
point is not simple. In addition, note that close to $T_c$, we have $dM/dT\propto-(T-T_c)^{\beta-1}$, and the magnetization 
exhibits a rapid decreasing that justifies the difficult to obtain measurements in the interval around to the 
critical temperature. For instance, the experimental values of $\beta$ are known for Rb$_2$MnF$_4$ 
($\beta\approx 0.16$)\cite{prb1.2211}, K$_2$MnF$_4$ ($\beta\approx0.15$)\cite{prb8.285}, 
MAMC ($\beta\approx 0.18$)\cite{pssb97.501}, and K$_2$NiF$_4$ ($\beta\approx 0.14$)\cite{prb1.2211,prb3.1736}. 

\begin{table}[h]
\centering
\begin{tabular}{l  l  l  l  l  l}
\hline\hline
				& Rb$_2$MnF$_4$ & K$_2$MnF$_4$		& MAMC 		& Rb$_2$MnCl$_4$ 	& K$_2$NiF$_4$	 \\
\hline
SCHA				& 0.155 			& 0.132				& 0.165 		& 0.121			 	& 0.187 	\\					
Exp.				& 0.16 $^a$		& 0.15$^b$			& 0.18$^c$	& -					& 0.138$^d$\\
\hline	
\end{tabular}
\label{table.beta}
\caption{Critical exponent $\beta$ for the magnetization power-law behavior $M(T)/M_s=B (1-T/T_c)^\beta$ obtained from SCHA 
and the respective experimental values.\\ $^a$ Phys. Rev. B 1, 2211 (1970)\\
$^b$  Phys. Rev. B 8, 304 (1973)\\ $^c$ Phys. Status Solidi (B) 97, 501 (1980)\\ 
$^d$ Phys. Rev. B 3, 1736 (1971)}
\end{table}
 
The aforementioned results were obtained utilizing the semiclassical (SCHA) approach. However, an alternative quantum approach 
can also be employed to determine the critical temperature. In this case, the same procedure is followed to obtain the values 
of $J$ and $\lambda$, which exhibit slight differences when determined using the quantum SCHA approach. 
As expected, the spectrum energy presents excellent agreement with the experimental data. However, the calculated critical 
temperature exceeds the experimental value. To address this discrepancy, we adopt a correction method developed by Ariosa and Beck, 
which is specifically designed to determine the BKT transition using the SCHA approach\cite{hpa65.499}. 
The authors argue that topological excitations show excessive energy, leading to an elevated critical temperature. 
The issue arises due to the replacement of the periodic cosine function 
$\cos\Delta\varphi$ by $\Delta\varphi^2$, which displays a single bump at $\Delta\varphi=0$. 
In our particular case, there are no vortex solutions; however, the excessive energy resulting from magnon excitations persists.
Indeed, consider the full probability function $P(x)=(1/Z)\exp[(\cos x)/t]$, where $t$ is a dimensionless parameter representing 
the reduced temperature, and the quadratic alternative $P_0(x)=(1/Z_0)\exp[(1-x^2/2)/t]$ ($Z$ and $Z_0$ are
normalization factors). Numeric integration for small values of $t$ reveals that 
$\langle\cos x\rangle\approx\langle(1-x^2/2)\rangle_0$. For example, when considering the interval $-\pi\leq x\leq \pi$, we
find that $\langle\cos x\rangle=0.95$ and $\langle(1-x^2/2)\rangle_0=0.9486$, when $t=0.1$. However, for $t=1$, the discrepancy 
increases, resulting in $\langle\cos x\rangle=0.509$ and $\langle(1-x^2/2)\rangle_0=0.446$. A similar observation can be made in the
SCHA, where $t=T/J_r S^2$ governs the average convergence. Observe that, in this case, we include the renormalization factor for
considering the real magnon energy in the calculation. In the semiclassical limit, $t\ll 1$, the SCHA method offers a 
more accurate approximation compared to the quantum limit, where the renormalization parameter $\rho$ is smaller 
than the semiclassical limit resulting in a larger $t$.

Therefore, following Ariosa and Beck, an improvement is achieved considering a three-bump probability function defined by 
$P(\Delta\varphi)=(1-p_\Delta)P_0(\Delta\varphi)+(p_\Delta/2)[P_0(\Delta\varphi-2\pi)+P_0(\Delta\varphi+2\pi)]$, where $P_0(\Delta\varphi)$ 
denotes the probability function associated with a single bump at $\Delta\varphi=0$ and $p$ represents the
probability for $\Delta\varphi$ to fall outside the interval $[-\pi,\pi]$. The angle average is then replaced by 
$\langle\Delta\varphi^2\rangle=\langle\Delta\varphi^2\rangle_0+4\pi^2p_\Delta$, where
\begin{equation}
p_\Delta=1-\textrm{erf}\left(\sqrt{\frac{(1+\lambda)\rho_\Delta\pi^2}{4t}}\right),
\end{equation}
and erf is the error function. Replacing the index $\Delta$ by $\Sigma$, we obtain a similar equation for $\langle\Sigma\varphi^2\rangle$.
The critical temperatures obtained through the quantum SCHA, endowed with the above correction, are presented in Table
(\ref{table.tc}). The obtained values of $T_c$ exhibit good agreement with experimental data. It is worth noting that, when 
utilizing the quantum approach, Rb$_2$MnCl$_4$ and MAMC demonstrate improved results when considering $S=5/2$ 
rather than the effective spin.

From an experimental standpoint, there is a limited availability of compounds with significant anisotropy. 
However, materials with large anisotropy can be studied by theoretical and computational simulation methods. 
Fig. (\ref{fig.tclambda}) shows the results
of SCHA, ($S\to\infty$) RPA analysis, and classical Monte Carlo simulation\cite{2dmater6.015028} 
regarding the relationship between critical temperature and anisotropy. 
The outcomes of (semiclassical) SCHA and RPA methods display close agreement, 
while the MC data yields lower values of $t_c$.
 
\begin{figure}[h]
\centering \epsfig{file=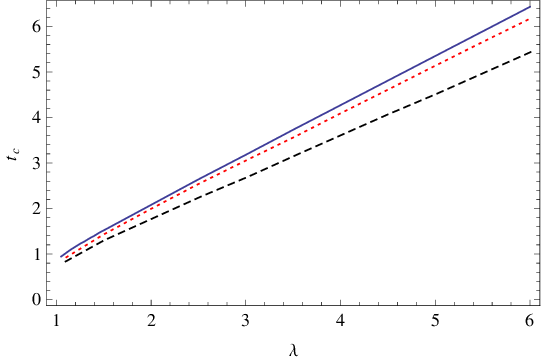,width=\linewidth}
\caption{The dependence on the anisotropy of the reduced critical temperature $t_c=T_c/JS^2$. The (blue) solid line is the SCHA result,
the (red) dotted line is from RPA analysis, and the (black) dashed line is data from Monte Carlo simulation.} 
\label{fig.tclambda}
\end{figure}

\section{Summary and Conclusions}
\label{sec.conclusions}
This paper explores the application of the Self-Consistent Harmonic Approximation (SCHA) to investigate 
two-dimensional magnetic models characterized by easy-axis anisotropy. While the SCHA formalism has 
been extensively employed to study phase transitions in various magnetic models, it has not been previously 
utilized for magnetic materials featuring easy-axis anisotropy. Thus, we have developed the SCHA approach, 
both in the semiclassical and quantum limits, for a general magnetic Hamiltonian that encompasses the easy-axis 
anisotropy. A comprehensive description is presented, encompassing essential information such as the energy 
spectrum, transition temperature, magnetization, and critical exponents for both ferromagnetic and 
antiferromagnetic models.

To validate the outcomes obtained using the SCHA method, we have applied it to describe five magnetic materials: 
Rb$_2$MnF$_4$, K$_2$MnF$_4$, MAMC, Rb$_2$MnCl$_4$, and K$_2$NiF$_4$. The first four materials exhibit a large spin value of $S=5/2$, 
while K$_2$NiF$_4$ has a spin of $S=1$. For the spin $S=5/2$ models, the semiclassical approach demonstrates excellent 
agreement with experimental data. However, the semiclassical SCHA results for K$_2$NiF$_4$ show only qualitative agreement. 
In contrast, the quantum SCHA method yields remarkable results for all five materials.

In summary, the SCHA formalism proves to be an outstanding tool for describing magnetic models 
featuring easy-axis anisotropy. We anticipate that this work can also be applied in the investigation of 
novel 2D Van der Waals systems and contribute to the understanding of magnetic devices in the field 
of spintronics, where easy-axis anisotropic magnetic materials are commonly used in spin-filters, for example.

\appendix
\section{Renormalization parameters} 
\label{appendix}
For determining the renormalization parameters, we compare the average $\langle \dot{S}_q^z\dot{S}_{-q}^z\rangle$
evaluated through the quadratic Hamiltonian, given by Eq. (\ref{eq.H0}), with the value obtained from the full Hamiltonian,
given by Eq. (\ref{eq.H}). The harmonic Hamiltonian provides the straightforward result
\begin{equation}
\langle\hbar^2 \dot{S}_q^z\dot{S}_{-q}^z\rangle_0=\frac{h_q^\varphi}{\beta},
\end{equation}
where $h_q^\varphi=zJS^2(\lambda\rho_x-\rho_y\gamma_q)$. 
In addition, using the full Hamiltonian, we can write the second average, in position space, as
\begin{equation}
\langle\hbar^2 S_i^z S_j^z\rangle=\frac{1}{Z}\int D\varphi DS^z\frac{1}{\beta}\frac{\partial^2 H}{\partial\varphi_i\partial\varphi_j}e^{-\beta H},
\end{equation}
where $Z$ is the partition function. The partial derivative is given by
\begin{IEEEeqnarray}{l}
\frac{\partial^2 H}{\partial\varphi_l\partial\varphi_{l^\prime}}=-J f_{ll^\prime}[(1+\lambda)\cos\Delta\varphi_{ll^\prime}+(1-\lambda)\cos\Sigma\varphi_{ll^\prime}]+\nonumber\\
+J\sum_j f_{lj}[(1+\lambda)\cos\Delta\varphi_{lj}-(1-\lambda)\cos\Sigma\varphi_{lj}]\delta_{ll^\prime},
\end{IEEEeqnarray}
where $l$ and $l^\prime$ represent nearest neighbor locations, and $f_{ll^\prime}$ is the same function
defined in Eq. (\ref{eq.H}). Therefore, the Fourier transform reads
\begin{IEEEeqnarray}{l}
\langle\hbar^2\dot{S}_q^z\dot{S}_{-q}^z\rangle=\frac{J}{2\beta N}\left\{-\sum_{\langle ll^\prime\rangle}[(1+\lambda)\langle f_{ll^\prime}\cos\Delta\varphi_{ll^\prime}\rangle+\right.\nonumber\\
+(1-\lambda)\langle f_{ll^\prime}\cos\Sigma\varphi_{ll^\prime}\rangle]e^{iq(l-l^\prime)}+\sum_{\langle lj\rangle}[(1+\lambda)\langle f_{lj}\cos\Delta\varphi_{lj}\rangle-\nonumber\\
\left.-(1-\lambda)\langle f_{lj}\cos\Sigma\varphi_{lj}\rangle]\right\},
\end{IEEEeqnarray}
where the factor of 2 was included to avoid double counting. 
Adopting independent bound averages, we achieve 
\begin{equation}
\langle f_{lj}\cos\Delta\varphi_{lj}\rangle\approx\left(S^2-\langle (S^z)^2\rangle_0\right)\exp\left(-\frac{\langle\Delta\varphi^2\rangle_0}{2}\right),
\end{equation}
and a similar equation for $\langle f_{lj}\cos\Sigma\varphi_{lj}\rangle$. Then, the previous equation becomes
\begin{IEEEeqnarray}{l}
\langle\hbar^2\dot{S}_q^z\dot{S}_{-q}^z\rangle=\frac{zJS^2}{2\beta}\{[(1+\lambda)\rho_\Delta-(1-\lambda)\rho_\Sigma]+\nonumber\\
+(1+\lambda)\rho_\Delta+(1-\lambda)\rho_\Sigma]\gamma_q\}.
\end{IEEEeqnarray}
The direct comparison with $\langle\hbar^2 \dot{S}_q^z\dot{S}_{-q}^z\rangle_0$ yields the self-consistent equations.
\bibliography{manuscript}
\end{document}